# Observations of Early Optical Afterglows

P. W. A. Roming[*] and K. O. Mason[†]

[*]Penn State University, Department of Astronomy & Astrophysics,
525 Davey Lab, University Park, PA 16802, USA
[†]Particle Physics and Astronomy Research Council,
Polaris House, North Star Ave, Swindon, Wilts SN2 1SZ, UK

**Abstract.** The Swift Ultra-Violet/Optical Telescope (UVOT) has performed extensive follow-up on 71 Swift Burst Alert Telescope triggered gamma-ray bursts (GRBs) in its first ten months of operations. In this paper, we discuss some of the UV and optical properties of UVOT detected afterglows such as XRF 050406, the bright GRB 050525A, the high redshift GRB 050730, the early flaring GRB 050801, and others. We also discuss some of the implications of why 75% of GRB afterglows observed by UVOT in less than one hour are "dark."

## INTRODUCTION

The Swift [1] Ultra-Violet/Optical Telescope (UVOT) [2] began observing gamma-ray burst (GRB) afterglows on January 24, 2005, after a three week on-orbit checkout. Between January 24 and December 9, 2005, the UVOT observed 71 GRB afterglows discovered by the Swift Burst Alert Telescope (BAT) [3], comprising 85% of the BAT observed sample. In addition, 60 of the 71 and 49 of the 71 afterglows were observed within less than one hour and five minutes, respectively, of the BAT trigger.

UVOT's speed of response, sensitivity, multiwavelength nature, and length of semi-uninterrupted observations[1] offers a unique opportunity to study the properties of early optical and UV afterglows. In addition, the Swift X-Ray Telescope (XRT) [4] positional data within the optical field of these afterglows allows us to be confident that no bright afterglows were missed due to confusion with nearby stars.

Below, we review some of the UV and optical properties of a sample of UVOT detected afterglows. In addition, we investigate some possible reasons why a large fraction of GRB afterglows observed by UVOT in less than one hour are "dark."

## DETECTED BURSTS

Between January 24 and December 9, 2005, the UVOT detected nineteen afterglows from GRBs discovered by the BAT. Table 1 identifies these UVOT detected afterglows as well as the broadband filters in which a detection was made. Redshifts, as determined by UVOT or ground-based data, are also provided. Below we

---
[1] Swift is in a low-Earth orbit and is therefore able to observe most objects continuously for no more than 40 minutes.

briefly describe the following GRBs: 050318, 050319, 050406, 050525A, 050730, 050801, 050802, and 051117A.

TABLE 1. UVOT Detected Afterglows

| GRB | UVW2 | UVM2 | UVW1 | U | B | V | White | z | Ref |
|---|---|---|---|---|---|---|---|---|---|
| 050318 | | | | X | X | X | | 1.44 | [5] |
| 050319 | | | | | X | X | | 3.24 | [6,7] |
| 050406 | | | | X | X | X | | 2.44 | [8] |
| 050416A | X | | | X | X | X | | 0.65 | [9,10,11,12] |
| 050525A | X | X | X | X | X | X | | 0.61 | [13,14] |
| 050603 | | | | | | X | | 2.82 | [15,16] |
| 050712 | | | | X | | X | | | [17,18] |
| 050726 | | | | | | X | | | [19,20] |
| 050730 | | | | | X | X | | 3.97 | [21,22,23] |
| 050801 | X | X | X | X | X | X | | <1.2 | [24] |
| 050802 | X | X | X | X | X | X | | <1.2 | [25,26] |
| 050815 | | | | | | X | | | [27,28] |
| 050820A | | | X | X | X | X | | 2.61 | [29,30] |
| 050824 | | X | X | X | X | X | | 0.83 | [31,32] |
| 050922C | | | X | X | X | X | | 2.07 | [33,34] |
| 051016B | X | | X | X | X | | | 0.94 | [35,36] |
| 051109A | | | X | X | X | X | | 2.35 | [37,38] |
| 051111 | | | X | X | X | X | | 1.55 | [39,40] |
| 051117A | | | | | X | X | X | | [41] |

## GRB 050318

GRB 050318 is the first burst for which UVOT detected an afterglow. The first UVOT detection occurred in the V-filter observation, which started 3230 seconds after the BAT trigger. The decay profile reveals very little fluctuation in the lightcurves, although the number of data points is admittedly small. A weighted mean of the decay slopes produces $\alpha = -0.94$. A combined UVOT and XRT spectrum provides a redshift of $z = 1.44$ [5].

## GRB 050319

Observations of GRB 050319 began much earlier, and were much more richly sampled, than GRB 050318. The first UVOT detection of GRB 050319 occurred in the V-filter observation, which started 90 seconds after the BAT trigger. The decay profile reveals potential fluctuations in both the V- and B-band lightcurves, particularly at early times. The best-fit power law is $\alpha = -0.57$ [6]. The XRT data reveals that a break in the fit to $\alpha = -1.14$ occurs at about 20,000 seconds after the trigger. No such break is evident in the late-time optical lightcurves. Based on an absorption line system in the spectrum of the GRB, a redshift of $z = 3.24$ is determined [7].

## GRB 050406

GRB 050406 is the first burst classified as an X-Ray Flash (XRF) with a detected UVOT afterglow; it was also the earliest optical detection of an XRF afterglow at the time. The first UVOT detection occurred in the V-filter observation, which started 88 seconds after the BAT trigger. The XRT lightcurve manifests flaring several hundred seconds after the burst, but no correlated flaring is seen in the UVOT lightcurves. The UVOT decay profile also does not follow the XRT's [8]. The X-ray lightcurve is well fitted by a broken power law of α = -1.58 at early times and α = -0.5 at later times [42]. The best power-law fit to the combined B- and U-band lightcurves produces α = -0.75. A UVOT and XRT broadband spectrum provides a redshift of $z = 2.44$ [8].

## GRB 050525A

GRB 050525A is the brightest UVOT afterglow detection to date and was well sampled in all the broadband color filters. The first detection occurred in the V-filter observation, which started 65 seconds after the BAT trigger. The maximum observed flux occurs 68 seconds after the burst, with $m_V = 12.86$. The decay profile for this burst is much more complex than the XRT lightcurve or any previous UVOT detected burst. The best fit to the lightcurve is a double power law, $α_1 = -1.56$ and $α_2 = -1.14$, with a decay index after a jet break of $α_j = -1.76$. The X-ray lightcurve is well fit by a broken power law of α = -1.2 at early times and α = -1.62 at later times [13]. Based on emission and absorption line systems in the spectrum of the GRB's host galaxy, a redshift of $z = 0.61$ is established [14].

## GRB 050730

The first UVOT detection of GRB 050730 occurred in the V-filter observation, which started 119 seconds after the BAT trigger [21]. The UVOT lightcurve reveals flaring in the first 1000 seconds with the peaks of the flares corresponding to the XRT flare peaks (see Figure 1). The best-fit optical profile, after the flaring, is a power law of α = -0.41, which is one of the shallowest decay profiles seen in UVOT lightcurves. The corresponding X-ray lightcurve is fit by a power law of α = -2.40, but variability in the lightcurve is a dominant feature. Based on a broad Lyα absorption line, thought to originate in the host of the GRB, a redshift of $z = 3.97$ is determined [22].

## GRB 050801

The first UVOT detection of GRB 050801 occurred in the V-filter, 52 seconds after the BAT trigger [24]. The X-ray lightcurve shows an initial decline and then a recovery. This is not unusual for the X-ray behaviour of bursts; however, what is unusual is that the optical curve follows it almost exactly (see Figure 2). The initial decline is even more significant in the UVOT data. After about 200 seconds both the X-ray and optical curves decline with α = -1.3. The X-ray lightcurve has a pretty good fit out to $10^5$ seconds, but there is excess optical emission with respect to the power law after about $10^4$ seconds. Using a combined XRT and UVOT SED, the best-fit redshift was determined to be $z = 1.45$.

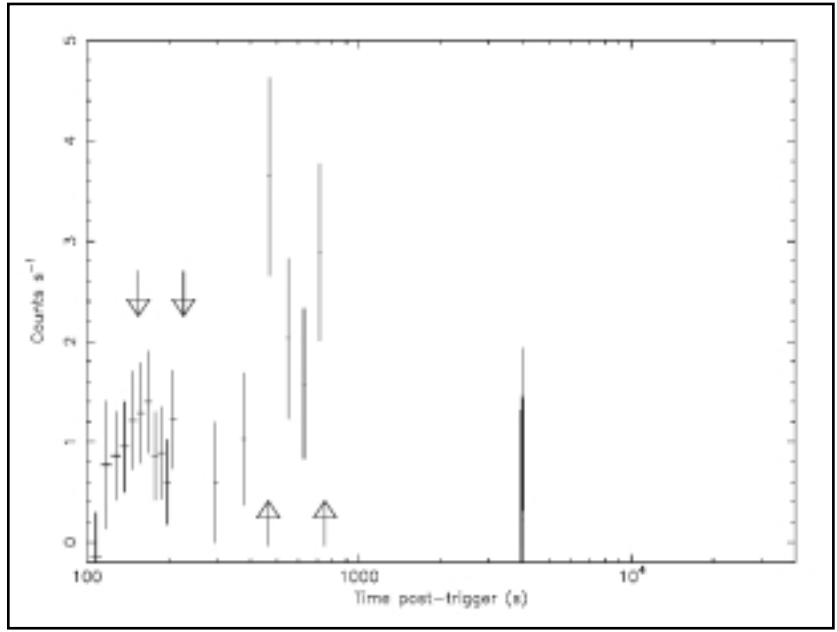

**FIGURE 1.** V-band lightcurve of GRB 050730. The arrows mark the times of the flares in the XRT lightcurve. Peaks in the V-band correspond with the locations of the X-ray flares.

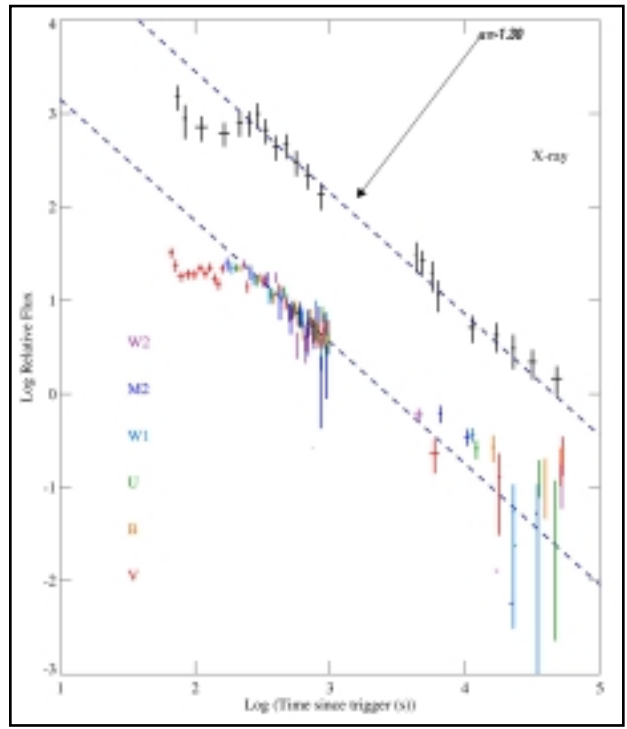

**FIGURE 2.** XRT and UVOT lightcurves of GRB 050801 plotted together in flux units on a log scale. The relative normalization between the X-ray and optical/UV curves is arbitrary for display purposes, but the same factor has been applied to the burst, so the separation does correspond to the relative X-ray to optical flux ratio. The UVOT data from all filters have been normalized to the V-band using data in the interval when the instrument was cycling rapidly between filters (10 second integration times) to determine the normalization factors. Filters distinguished by color.

## GRB 050802

The first UVOT detection of GRB 050802 occurred in the V-filter, 286 seconds after the BAT trigger [25]. As with GRB 050801, there is some initial variability in the X-ray lightcurve which is rising followed by a small re-brightening at about 1000 seconds. There is no strong evidence for similar behaviour in the UVOT lightcurve, but the error bars are larger than GRB050801 (see Figure 3). The overall decline is much flatter than 050801, with $\alpha = -0.9$. There is a break in the X-ray band at about $10^4$ seconds to $\alpha = -1.6$, but no corresponding break in the UV/optical. Based on several absorption features, a redshift of $z = 1.71$ was determined [43].

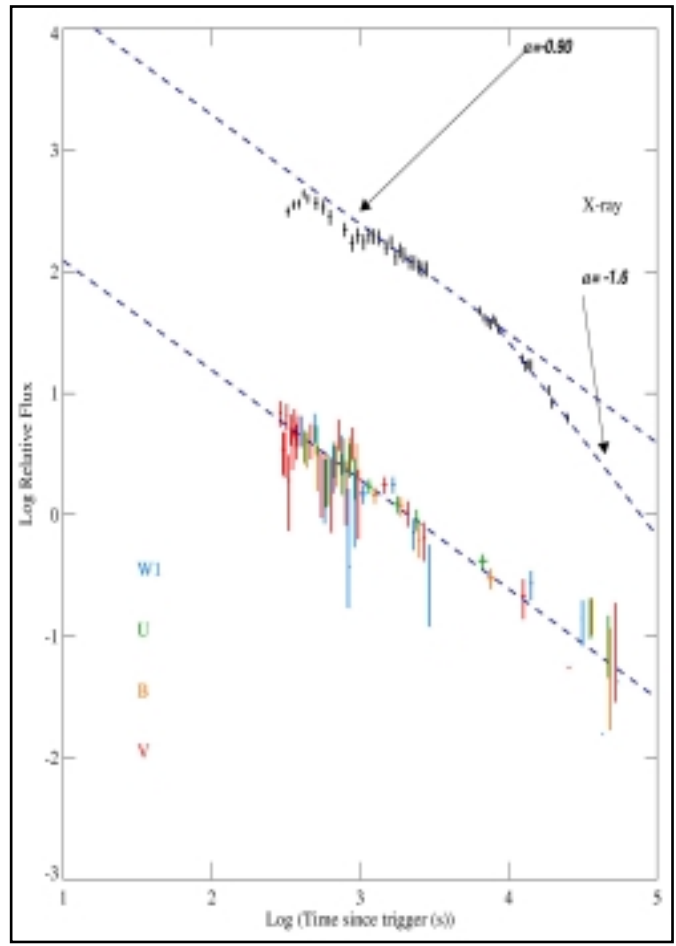

**FIGURE 3.** XRT and UVOT lightcurves of GRB 050802 plotted using the same technique as was used for GRB 050801 (see Figure 2).

## GRB 051117A

The UVOT White-light filter was inserted into the automated observing sequence after the first nine months of operation. The White-light filter is effectively a clear filter, and the response curve is essentially the wavelength-dependant response curve of the detector. These observations are thus similar to the unfiltered observations made by ROTSE [44] and should be more sensitive than filtered observations for low-redshift bursts. The first UVOT detection of GRB 051117A occurred in the V-filter, 111 seconds after the BAT trigger [41]. However, later detections of the afterglow were made in the White-light filter revealing that the source was indeed fading (see Figure 4). Because the observations were made in the White-light filter, no redshift determination could be made.

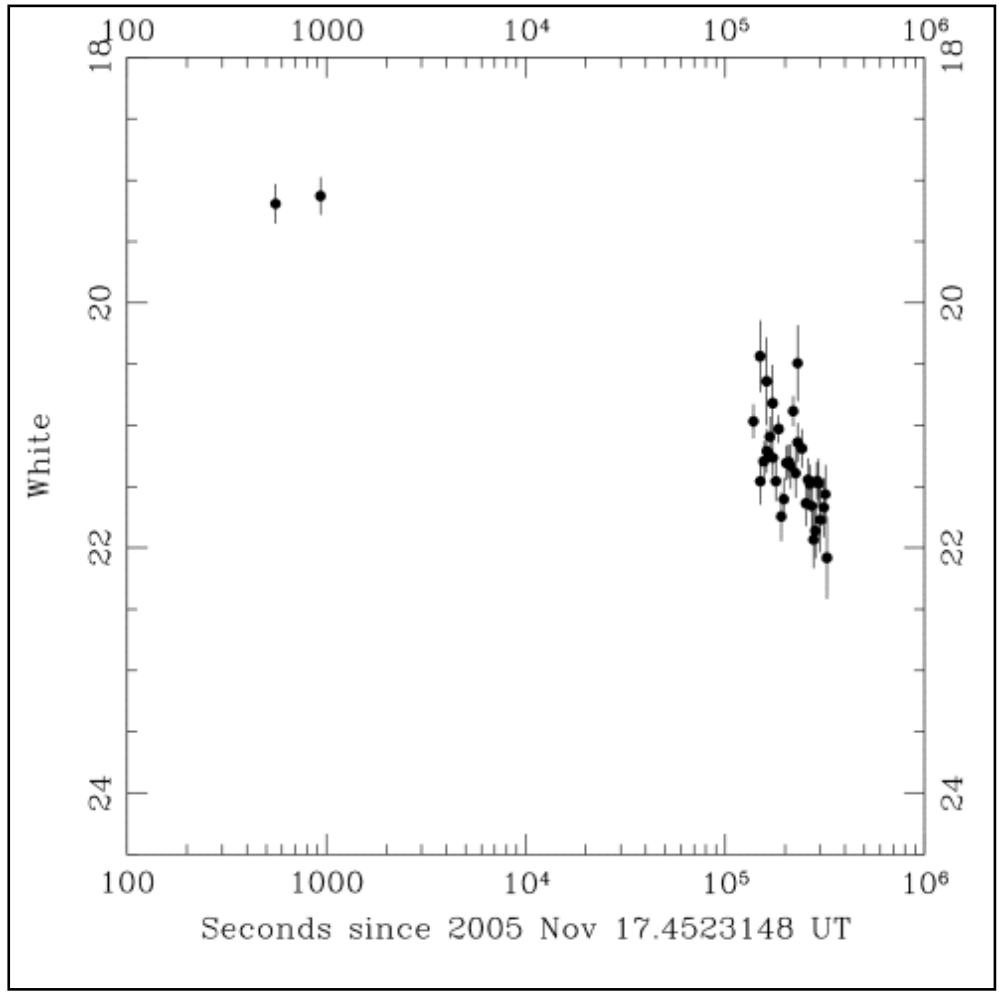

**FIGURE 4.** White-light observation of GRB 051117A. The source is clearly detected and is fading. The Figure shows all the White-light observations up to approximately 3.5 days after the BAT trigger that are above the 3-sigma detection limit.

## General Properties of UVOT Detected Afterglows

Because the number of UVOT detected afterglows is still small, it is difficult to quantity the sample properties of these early afterglows at this time. However, we provide the distribution of the redshift (see Figure 5) as well as the temporal decay indices (see Figure 6) for the UVOT detected afterglows. From the figures, it appears that the normal distribution of UVOT bursts is in the $0 < z < 3$ range and the temporal decay indices are weighted to the 0.5-1.0 range. Qualitatively, we can describe a range of properties seen in UVOT detected afterglows, as follows:

- there is no UV/optical flaring although there is X-ray flaring,
- there is UV/optical flaring as well as X-ray flaring,
- fluctuations are seen in shallow decay UV/optical lightcurves,
- fluctuations are seen in steep decay UV/optical lightcurves,
- UV/optical lightcurves follow the X-ray lightcurves,
- UV/optical lightcurves do not follow the X-ray lightcurves,
- steep decay profiles are consistent with the standard fireball model,
- and, shallow decay profiles are difficult to reconcile with the standard fireball model.

From the description above, it can be seen that the optical lightcurves have large range of properties, pointing to different circumburst environments. Again, it is emphasized, that the sample size is still small. As the sample grows it may become apparent that a particular feature or set of features will dominate the properties of the early afterglows of GRBs.

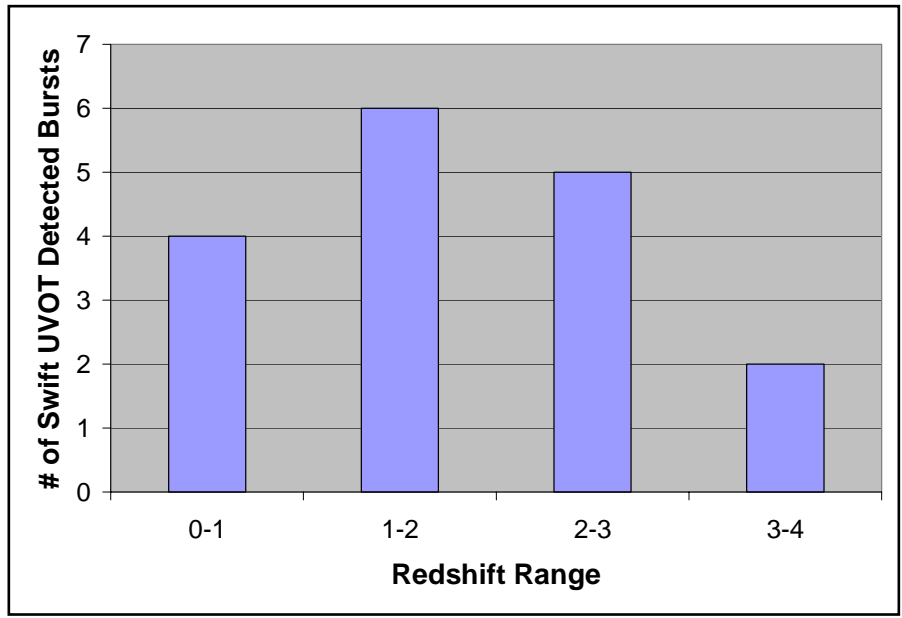

**FIGURE 5.** Distribution of the redshift range for UVOT detected afterglows.

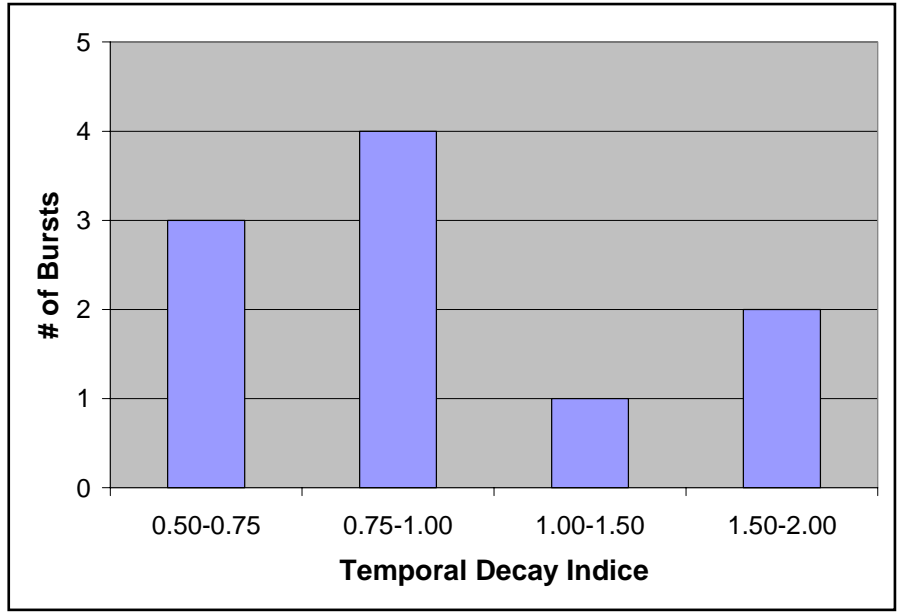

**FIGURE 6.** Distribution of the temporal decay indices for UVOT detected afterglows.

## "DARK" BURSTS

UVOT observed 60 bursts in less than an hour (and in 49 cases in less than five minutes) after the BAT trigger with no detections being made in 45 instances. Of the 45 non-detections with UVOT, 13 *were* detections by ground-based observers via the GCN [45]. Most of these reported detections were in bands red-ward of the UVOT's, perhaps indicative of a high redshift such that the Lyman edge is beyond the UVOT range. It may be that some of the remaining 32 non-detections could be due to them being inaccessibility to ground-based R- and IR-band observations in that part of the sky. However, even if the number of ground detections is doubled, approximately 25% of bursts would still be classified as "dark."

Qualitative definitions have been used to describe dark bursts such as: no afterglow detection in the optical or NIR, and the optical counterpart is initially faint and declines too rapidly to be detected. More recent work has tried to quantify the definition of dark bursts in order to remove the dependence on instrument sensitivity. The optical–to–X-ray spectral index method ($\beta_{OX}$) [46] is a quick method to determine if Swift bursts are dark. If $\beta_{OX} < 0.50$, then the burst is classified as dark. If $0.50 < \beta_{OX} < 0.55$, then the burst is classified as potentially dark. A more rigorous treatment is the upper limit on the afterglow flux (ULAF) method [47]. The ULAF method uses the temporal ($\delta$) and spectral ($\beta$) indices to determine the minimum and maximum value for the electron index (p). It examines eight cases of the standard afterglow model and extrapolates the maximum & minimum X-ray flux to the optical epoch. Optical flux that is below the minimum extrapolated X-ray flux is considered dark.

We have selected 19 bursts – which include some that are detected and others that are not detected by the UVOT – as a test of the $\beta_{OX}$ and ULAF methods. Using the ULAF method, only one of the 19, GRB 050219B, is classified as dark. Using the $\beta_{OX}$

method, GRBs 050315, 050319, 050326, 050412, and 050505 are classified as dark [48]. It is noted that GRB 050319 was detected by UVOT [6], and that GRBs 050315 [49] and 050505 [50] were detected by ground-based observations.

Although many of the bursts in our sample are not classified as dark in the ULAF or $\beta_{OX}$ classification schemes, it is still unclear why the early optical afterglows are suppressed. Both the $\beta_{OX}$ and ULAF methods assume that GRBs follow the standard fireball model. However, recent work [51,52] suggests that the standard fireball model is not valid in the first few hours after many bursts. What then are the explanations for these dark bursts at early times?

Conventional reasons put forth for dark bursts include: circumburst absorption [53,54], low-density environment [55,56,57], intrinsic faintness [53,54], rapid temporal decay [56], Ly-α blanketing and absorption due to high redshift [54,58], and extinction by dust [59]. For UVOT observed afterglows in which there was no detection, the standard explanations for darkness can be invoked in many cases. However, for bursts such as GRBs 050223, 050421, & 050422, the conventional explanations can't explain the suppression of the afterglow.

Recent work suggests that the suppression of these early GRB afterglows can be explained by high gamma-ray efficiencies which are consistent with highly magnetized outflows [48]. By comparing the X-ray afterglow flux at one hour ($F_{x,1}$) to the prompt γ-ray fluence (Sγ) a gamma-ray efficiency can be defined. As the sample of "dark" bursts increases, we expect that our understanding of the suppression mechanisms will also continue to grow.

## CONCLUSIONS

In less than one year of operation, the UVOT, in conjunction with other ground based telescopes, has more fully mapped the early time optical parameter space of GRB afterglows than previous telescopes or missions. Currently, the UVOT optical lightcurves denote a large array of properties which indicates a range of circumburst environments. Additionally, Swift data on dark bursts points to a newly proposed mechanism than the standard explanations previously presented. As the sample continues to grow, we anticipate that a specific trend or trends will reveal themselves.

## ACKNOWLEDGMENTS

This work is supported at Penn State by NASA contract NAS5-00136. We gratefully acknowledge the contributions from members of the Swift UVOT team at PSU, MSSL, NASA/GSFC, and our subcontractors.